\documentclass{IEEEtran4PSCC}
   \usepackage[dvips]{graphicx}
  \graphicspath{{./}}
\newcommand{\keno}{\hspace{7pt}}
\usepackage{xcolor}
\usepackage{mathtools}
\usepackage{comment}
\usepackage{svg}
\usepackage{subcaption}
\usepackage{multirow}
\usepackage{url}
\usepackage{algorithm2e}

\newcommand{\hl}[1]{{\color{black}#1}} 

\usepackage{amsfonts}
\newcommand{\maxpd}{\overline{\mathrm{\mathbf{p}}}_{d}}
\newcommand{\minpd}{\underline{\mathrm{\mathbf{p}}}_{d}}
\newcommand{\pgslack}{(\hat{\boldsymbol{p}}_g)^{slack}}
\newcommand{\weightk}{{\mathbf{W}}_{k+1}}
\newcommand{\weightK}{{\mathbf{W}}_{K+1}}
\newcommand{\biask}{{\mathbf{b}}_{k+1}}
\newcommand{\biasK}{{\mathbf{b}}_{K+1}}
\newcommand{\pd}{{\boldsymbol{p}}_{d}}

\newcommand{\setbuses}{\mathbf{B}}

\renewcommand{\baselinestretch}{0.96}
\makeatletter
\let\old@ps@headings\ps@headings
\let\old@ps@IEEEtitlepagestyle\ps@IEEEtitlepagestyle
\def\psccfooter#1{%
    \def\ps@headings{%
        \old@ps@headings%
        \def\@oddfoot{\strut\hfill#1\hfill\strut}%
        \def\@evenfoot{\strut\hfill#1\hfill\strut}%
    }%
    \def\ps@IEEEtitlepagestyle{%
        \old@ps@IEEEtitlepagestyle%
        \def\@oddfoot{\strut\hfill#1\hfill\strut}%
        \def\@evenfoot{\strut\hfill#1\hfill\strut}%
    }%
    \ps@headings%
}
\makeatother

\psccfooter{%
        \parbox{\textwidth}{\hrulefill \\ \small{23rd Power Systems Computation Conference} \hfill \begin{minipage}{0.2\textwidth}\centering \vspace*{3pt} \includegraphics[scale=0.06]{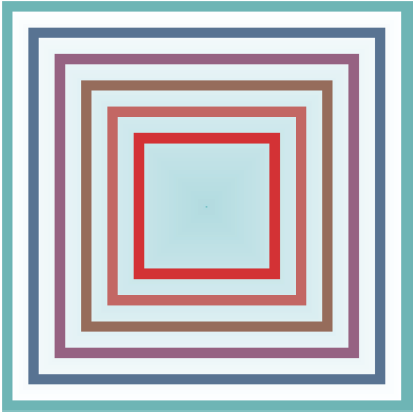}\\\small{PSCC 2024} \end{minipage} \hfill \small{Paris, France --- June 4 -- 7, 2024}}%
}
\usepackage{amsmath}

\begin{document}
%
\title{A hybrid Quantum-Classical Algorithm for Mixed-Integer Optimization in Power Systems}

\author{
\IEEEauthorblockN{Petros Ellinas, Samuel Chevalier, Spyros Chatzivasileiadis,}
\IEEEauthorblockA{Department of Wind and Energy Systems, 
 Technical University of Denmark (DTU)\\
 Elektrovej, 2800 Kgs. Lyngby, Denmark\\
}
}


\maketitle

\begin{abstract}
Mixed Integer Linear Programming (MILP) can be considered the backbone of the modern power system optimization process, with a large application spectrum, from Unit Commitment and Optimal Transmission Switching to verifying Neural Networks for power system applications. The main issue of these formulations is the computational complexity of the solution algorithms, as they are considered NP-Hard problems. Quantum computing \hl{has been tested as a potential solution towards reducing the computational burden imposed by these problems, providing promising results, motivating the can be used to speedup the solution of MILPs}. In this work, we present a general framework for solving power system optimization problems with a Quantum Computer (QC),  which leverages mathematical tools and QCs' sampling ability to provide accelerated solutions. Our guiding applications are the optimal transmission switching and the verification of neural networks trained to solve a DC Optimal Power Flow. Specifically, using an accelerated version of Benders Decomposition , we split a given MILP into an Integer Master Problem and a linear Subproblem and solve it through a hybrid ``quantum-classical'' approach, getting the best of both worlds.  We provide 2 use cases, and benchmark the developed framework against other classical and hybrid methodologies, to demonstrate the opportunities and challenges of hybrid quantum-classical algorithms for power system mixed integer optimization problems.
\end{abstract}


\thanksto{\noindent Submitted to the 23rd Power Systems Computation Conference (PSCC 2024). This work is supported by the ERC Project VeriPhIED, funded by the
European Research Council, Grant Agreement No: 949899.}

\section{Introduction}
Mixed-integer linear programs (MILPs) are pivotal in the realm of power systems engineering. These mathematical optimization methods, dubbed as such due to their inclusion of both continuous and discrete variables, are harnessed for the modeling and resolution of a diverse array of issues related to the operation, planning, and control of electrical power systems. For example, fundamental challenges such as unit commitment \cite{unit_commitment}, investment planning \cite{plannin_comb}, and optimal transmission switching (OTS) \cite{hinneck2021optimal} are cast as MILPs. Furthermore, there is a growing trend toward utilizing additional tools, such as Neural Network constraint-violation verification \cite{venzke2020learning}, which can also be framed as MILPs. 

Nevertheless, the integer aspect introduces a combinatorial element to the problem, as the solver must traverse various combinations of discrete choices. With an increase in the number of discrete variables and their potential values, the solution space grows exponentially, classifying the problem of finding the global solution of MILP models within the NP-Hard problem category. 

Quantum computing is a promising direction to deliver a solution to the high computational complexity of MILPs. Recently, authors in \hl{\cite{tasseff2022emerging},\cite{brown2024copositive}} have provided compelling evidence to support the notion that QCs can currently offer valuable capabilities, even with their inherent noise at the moment, and prior to the advent of fault-tolerant systems. This includes the currently available processors. Considering the rapid progress in the quantum computing field, we expect that these capabilities will increase substantially when effective fault-tolerant systems for QCs develop in the near future. Therefore, it is necessary to explore this track, as it can be a methodology that would help us solve a set of MILP power system problems that have so far been impossible to solve \hl{in an acceptable time window with reasonable computational recourses}, and operate more efficient, safe, and resilient power grids. 

Quantum computing has been investigated for the solution of MILPs, for instance, in \cite{zhao_quantum_milp} where the authors propose a Benders decomposition approach to solve a general form of MILPs, using a QC. However, according to \cite{PATERAKIS2023108161} there are numerous challenges resulting from this method, such as the choice of inferior feasibility and optimality cuts and slow convergence towards the end of the algorithm. In \cite{PATERAKIS2023108161}, a multi-cut Benders decomposition is presented, during which the Quantum Processing Unit (QPU) is specifically allocated to execute a binary optimization subroutine focused on selecting the cuts that should be incorporated into the Master Problem (MP) \hl{to minimize the problem's size and accelerate the solution}. However, this is done while both the MP and the Subproblems (SPs) are solved through classical methods, which slows the algorithm down significantly. 

In the current literature, the exploration of quantum algorithms' application to power systems is still limited \cite{GOLESTAN2023584}. Among them, Ref. \cite{saevarsson2022quantum} presented the first successful solution of an AC power flow in a real QC. They successfully solved it for 3 and 5 bus systems, and they discussed scalability and noise issues of the current "noisy-era" QCs. \hl{Before that, Ref \cite{feng2021quantum} proposed an algorithm to solve a power flow with quantum computing but only demonstrated in simulated quantum simulations.} Ref. \cite{zhou2022noise} put forward the idea of combining machine learning and quantum computing to potentially tackle security and reliability issues in large-scale power grid systems. \hl{Furthermore, in \cite{jones2020computational} authors use Quantum Annealing to solve a simple version of optimal phasor measurement unit placement, which is a crucial problem for the next-generation power system design. They showed that in certain cases, the Quantum Annealing solver surpassed the classical solver CPLEX in performance.}

\hl{Many studies have been focused in solving the Unit Commitment problem. In particular, \cite{unit_commitment_lit} proposed a distributed algorithm for solving unit commitment. Furthermore, \cite{FENG2022108386} proposed quantum-circuit-based algorithms for microgrid state estimation. Following the work of \cite{Gambella2020MultiblockAH}, authors of \cite{Mahroo_2022} have proposed a three-block alternating direction method, to solve the unit commitment problem.}

In this work, we provide a framework for transforming MILP models of power system problems to a "quantum-classical solvable form". In contrast to alternative approaches, our methodology incorporates mathematical tools or harnesses the sampling capabilities of quantum solvers to expedite the solution process. \hl{Specifically, in quantum computing, measurements collapse the superposition of qubits' states to a specific outcome. The probabilities of different outcomes are determined by the amplitudes of the quantum states. Repeated measurements allow for the sampling of these outcomes according to their probabilities. Similarly, at the end of the annealing process, the quantum annealer samples from the final quantum state to obtain solutions to the optimization problem.} The resulting combinatorial problems are solved solely in a QPU. \hl{The derived approximated solution, in some cases, has been empirically proven to converge faster to the optimal solution than the classical branch and bound method.}  Additionally, we offer a comprehensive Python code in the form of a tutorial available on our GitHub repository (\cite{github}), which instructs users on how to convert any Mixed-Integer Linear Program (MILP) into the specified format.  We benchmark this approach on OTS \cite{hinneck2021optimal} and NN verification for DC-OPF \cite{venzke2020learning}, using both of the two prevailing quantum approaches: quantum annealing and quantum circuits.


Summarizing, the contributions of this paper are:
\begin{itemize}
    \item We provide a framework for transforming MILP models of power system problems to a "quantum-classical solvable form", providing two different acceleration techniques. As part of the contributions of this paper, a comprehensive Python code, in the form of a tutorial, on how to solve any MILP problem in this framework is made publicly available in \cite{github}. 
    \item We provide extensive numerical results for large enough instances that arrive at the limits of the capacity constraints of the current available QCs and we discuss the benefits of each method and the challenges that need to be addressed to exploit the future quantum advantage.
\end{itemize}

This paper is organized as follows. Section~\ref{prob_form} covers Benders decomposition theory, its enhancements, and the conversion of the MP to a quantum solvable Quadratic Unconstrained Optimization problem (QUBO). In Section~\ref{benchmarks}, we detail the optimization models for benchmarking, while Section~\ref{numericals} showcases the numerical experiments evaluating the framework's performance. Section~\ref{sec:discussion} discusses the results and current challenges, and Section~\ref{sec:conclusion} concludes.

\section{Problem Formulation} \label{prob_form}
A variety of power system problems can be formulated in the form:
\begin{equation} \label{first_form}
\begin{aligned}
\max_{\boldsymbol{z},\boldsymbol{y}} \quad &\mathbf{i}^{T} \boldsymbol{z} + \mathbf{c}^{T} \boldsymbol{y} \\
    s.t. \quad &\mathbf{A} \boldsymbol{z} + \mathbf{B} \boldsymbol{y} \leq \mathbf{b} \\
     & {\boldsymbol{z}} \in \mathbb{Z}, {\boldsymbol{z}} \in \{0,1\}^n, \boldsymbol{y} \in \mathbb{R}^{p}_{+}
\end{aligned} 
\end{equation}
where $\mathbf{i}^{T}$ and $\mathbf{c}^{T}$ are the coefficient matrices of integer and continuous variables in the objective function, and $\mathbf{A}$ and $\mathbf{B}$ correspond to coefficient matrices of integer and continuous variables in the constraints, respectively. Moreover, $\mathbf{b}$ is a vector of real numbers.  These models are a special form of a bigger umbrella of the MILP models because they consist of binary variables $\mathbf{z}$ (instead of general integer variables) and continuous variables $\mathbf{y}$. 

To exploit the speedup of quantum algorithms \cite{exponentialsomma2012quantum,exponentialzhou2020quantum} in solving combinatorial optimization problems, which are integer programs with binary variables, we should separate the aforementioned MILP models into an integer program and a continuous program. This can be done using Benders Decomposition (BD) \cite{benders}. The main idea of Benders is that for each feasible $\boldsymbol{z}$, the optimal $\boldsymbol{y}$ is determined via a linear program. Treating $\boldsymbol{y}$ as a function of $\boldsymbol{z}$, we use a scalar variable to represent the best $\boldsymbol{y}$ choice for a given $\boldsymbol{z}$, replacing its contribution to the objective. Initially, a coarse estimation of $\boldsymbol{y}$ is employed, followed by a sequence of dual solutions to tighten the approximation. This process converts the original MILP problem into a simpler MILP Master Problem and a linear Subproblem. 

According to the above, we can first fix the binary variables in \eqref{first_form}, transforming them to the vector $\boldsymbol{z}^{(n)}$. Therefore we can write the primal form of the SubProblem as a linear program:
\begin{align}
    \max_{\boldsymbol{y}} \quad &\mathbf{i}^{T} \boldsymbol{z}^{(n)} + \mathbf{c}^{T} \boldsymbol{y} \tag{SP} \label{sub_problem}\\
    s.t. \quad &\mathbf{B} \boldsymbol{y} \leq \mathbf{b} - \mathbf{A} \boldsymbol{z}^{(n)}  \quad :\boldsymbol{\lambda} \label{linear_constr}
\end{align}
where $(n)$ index denotes the current iteration number. As \eqref{sub_problem} is a convex linear program consisting only of continuous decision variables, a strong duality condition holds. Consequently we can define $\boldsymbol{\lambda}$ as the vector of dual variables associated to constraints \eqref{linear_constr} and so we can express its dual form as:
\begin{align}
      x(\boldsymbol{z}) = \min_{\boldsymbol{\lambda}} \quad &\boldsymbol{\lambda} (\mathbf{b}- \mathbf{A} \boldsymbol{z}^{(n)})^{T} \tag{Dual-SP} \label{dualsub_sub_problem}\\
    s.t. \quad &\mathbf{B} \boldsymbol{\lambda} \geq \mathbf{c}  \notag \\
               & \boldsymbol{\lambda} \geq 0 \notag
    \end{align}
At this point, we define extreme points set $\mathbf{T}$ and extreme rays set $\mathbf{K}$. We denote extreme point $t$ as $\boldsymbol{p}^{t}$ and extreme ray $k$ as $\boldsymbol{r}^{k}$ and they belong to the polyhedron:
\begin{equation}
    Q=\{\boldsymbol{\lambda} \in {\mathbb{R}_{+}^{m}} \keno | \keno \mathbf{B} \boldsymbol{\lambda} \geq \mathbf{c} \}
\end{equation}
Extreme rays and extreme points can be acquired from \eqref{dualsub_sub_problem}'s decision variables. Specifically, the \eqref{dualsub_sub_problem} is unbounded and $x(\mathbf{z})=-\infty$, then $\boldsymbol{\lambda}$ expresses a dual ray $\boldsymbol{r}^{k^{'}}$, rendering \eqref{sub_problem} infeasible. This means that $x$ does not allow a feasible solution to \eqref{first_form}, and that helps us to introduce a new feasibility cut in the MP as:
\begin{equation}
    (\mathbf{b}- \mathbf{A} \boldsymbol{z}) \boldsymbol{r}^{k^{'}} \geq 0 \quad k^{'} \in K
\end{equation}
However when \eqref{dualsub_sub_problem} is feasible, then dual variables $\boldsymbol{\lambda}$ denote an extreme point $\boldsymbol{p}^{t^{'}}$, shaping an optimality cut as:
\begin{equation}
    (\mathbf{b}- \mathbf{A} \boldsymbol{z}) \boldsymbol{p}^{t^{'}} \geq s \quad t^{'} \in T
\end{equation}
where s is a scalar value denoting the contribution of continuous variables to the MP:

\begin{align}
    \max_{\boldsymbol{z}} \quad & \mathbf{i}^{T} \boldsymbol{z} + s \tag{MP} \label{master}\\
 \quad s.t. \quad & \boldsymbol{p}^{t}(\mathbf{b} - \mathbf{A} \boldsymbol{z}) \geq s \quad \text{for } t \in T \notag \\
      \quad & \boldsymbol{r}^{k} (\mathbf{b} - \mathbf{A} \boldsymbol{z}) \geq 0 \quad \text{for } k \in K \notag
\end{align}

The algorithm iteratively reduces the optimality gap. The process is interrupted if the optimality gap is lower than a small positive number $\epsilon$. It has been shown that Benders Decomposition converges to the global optimum of the initial MILP problem \cite{PATERAKIS2023108161}. 
\subsection{Improvements}
\subsubsection{Optimization Method I}
It has been reported that in some cases Benders Decomposition can have slow convergence times. However, as pointed out in references \cite{magnanti1981accelerating,acc2,PAPADAKOS2008444}, there are multiple strategies, from the classical mathematical programming perspective, to enhance Benders decomposition. Consequently, we modify the conventional BD algorithm to tackle the slow convergence problems, by guiding the choice of binary variables and selecting efficient cuts.

Smart cuts or Pareto-optimal cuts selection were investigated in \cite{magnanti1981accelerating}. The authors of the aforementioned paper first introduced the term "domination" for cuts. Specifically a cut $l^{'}(\mathbf{b} - \mathbf{A} \boldsymbol{z})$ dominating the cut $l^{''}(\mathbf{b} - \mathbf{A} \boldsymbol{z})$ when:
\begin{equation}
    l^{'}(\mathbf{b} - \mathbf{A} \boldsymbol{z}) \geq l^{''}(\mathbf{b} - \mathbf{A} \boldsymbol{z}) \quad \forall \quad \boldsymbol{z} \in Z
\end{equation}
for all $\boldsymbol{z}$ for a strict inequality for at least one solution $\boldsymbol{z}$. Therefore they denote the pareto-optimal cut as the cut that does not get dominated by any other cuts.

Furthermore in \cite{magnanti1981accelerating}, the definition of the core point was introduced and they connected it to the selection of Pareto optimal cuts. The core point of $Z$ (denoted as $\overline{\boldsymbol{z}}^{(n)}$) is any point from the set $Z$ that lies within the relative interior of the convex hull $Y_c$ of $Y$. Finally, \cite{PAPADAKOS2008444}, exploited the connection between core points and Pareto optimal cuts, proposing an independent optimization problem to generate Pareto optimal cuts as: 
\begin{align}
    \min_{\boldsymbol{\lambda}} \quad &\boldsymbol{\lambda} (\mathbf{b}- \mathbf{A} \overline{\boldsymbol{z}}^{(n)})^{T} \tag{Pareto-SP} \label{additional}\\
    s.t. \quad &\mathbf{B} \boldsymbol{\lambda} \geq \mathbf{c}  \notag \\
               & \boldsymbol{\lambda} \geq 0 \notag
\end{align}
where $\overline{\boldsymbol{z}}^{(n)}$ is a core point of $Z$. The optimal solution in the proposed problem is Pareto optimal.

Still, according to \cite{MERCIER20051451,mercier2007integrated}, it is very hard to find a core point $\overline{\boldsymbol{z}}^{(n)}$. To tackle that, \cite{PAPADAKOS2008444} used the following core point approximation technique:
\begin{equation}
    \overline{\boldsymbol{z}}^{(n)} = \frac{1}{2} \overline{\boldsymbol{z}}^{(n-1)} + \frac{1}{2} {\boldsymbol{z}}^{(n-1)}
\end{equation}
with a random $\overline{\boldsymbol{z}}^{(0)}$. With this we can approximate a core point gradually, as in every iteration, we get a point between two solutions, which gradually approaches a core point.

Another problem of BD is the tailing effect \cite{acc2}. Specifically, in \cite{acc2}, it has been observed that algorithms present an unstable behavior when they approximate the solution. Therefore, they proposed to add an extra regularizing term to the objective function. This term is the Hamming distance between $\overline{\boldsymbol{z}}^{(n)}$ and $\overline{\boldsymbol{z}}^{(n-1)}$. Consequently, we can write the resulting MP as:
\begin{align}
    \max_{\boldsymbol{z}} \quad & \mathbf{i}^{T} \boldsymbol{z} + s + ||{\boldsymbol{z}}-{\boldsymbol{z}}^{(n-1)}|| \tag{MP-Opt} \label{master_optimized}\\
 \quad s.t. \quad & \boldsymbol{p}^{t}(\mathbf{b} - \mathbf{A} \boldsymbol{z}) \geq s \quad \text{for } t \in T \notag \\
      \quad & \boldsymbol{r}^{k}(\mathbf{b} - \mathbf{A} \boldsymbol{z}) \geq 0 \quad \text{for } k \in K \notag
\end{align}
A benefit of this method is that we can remove previous cuts if the algorithm has to execute many iterations \cite{acc2}.  This can reduce the memory alongside the qubits requirements of the algorithm. The whole algorithm with optimization method I can be described by Algorithm \ref{alg:opt_group_1}.

\RestyleAlgo{ruled}

\begin{algorithm} 
\caption{Optimization \hl{Method} I} \label{alg:opt_group_1}
\KwData{$\text{Problem } \eqref{first_form}$}
\KwResult{$\boldsymbol{z}^{*},\boldsymbol{y}^{*}$}
$n \leftarrow 1$\;
$\boldsymbol{z},\overline{\boldsymbol{z}}^{(0)} \leftarrow 0$\;
\While{$|UB-LB|>\epsilon$}{
    $\text{Solve } \eqref{dualsub_sub_problem} \text{ and get } x^{(n)}({\boldsymbol{z}})$\;
     $LB \leftarrow x^{(n)}_{\boldsymbol{z}} $\;
     $\overline{\boldsymbol{z}}^{(n)} = \frac{1}{2} \overline{\boldsymbol{z}}^{(n-1)} + \frac{1}{2} {\boldsymbol{z}}^{(n-1)}$\;
 $\text{Solve } \eqref{additional} \text{ and get } x(\overline{\boldsymbol{z}}^{(n)}), \boldsymbol{\lambda} $\;
     \eIf{$x(\overline{\boldsymbol{z}}^{(n)}))=-\infty$}{
        $\text{Add extreme ray } K \cup \{k\}$\;
      }{
        $\text{Add extreme point } T \cup \{t\}$\;
      }
     $\text{Solve } \eqref{master_optimized} \text{ and get } s(\boldsymbol{p}),\boldsymbol{z}^{(n)}$\;
    $UB \leftarrow s(\boldsymbol{p}) $\;
    $n \leftarrow n+1$\;
}
\end{algorithm}


\subsubsection{Optimization Method II}
This optimization methodology exploits the ability of quantum algorithms to provide multiple solutions at the same time, with a technique called sampling. During our experiments, we found that the exchange of data between QPU and CPU is very inefficient. Consequently, to minimize the access time to the QPU, in every iteration we choose to run the \eqref{dualsub_sub_problem} for the best $R$ solutions that the QPU provides. Therefore, Algorithm \ref{alg:opt_group_2} leverages the ability of QCs to provide multiple solutions, to accelerate the solution of MILPs.

\subsection{QUBO transformation}
Despite the two reformulations in the linear SPs, the MP resulting from Benders Decomposition cannot be directly solved by a QC, as it is still a MILP, \hl{as it contains the scalar variables}. \hl{As provided evidence e.g. \cite{chang2023quantum} show that QC can solve certain combinatorial problems faster than classical computers \cite{chang2023quantum}}, we try to transform the \eqref{master_optimized} to a "quantum solvable" form: that is a QUBO. \hl{Note that Higher-Order Binary Optimization (HOBO) is applicable as it can be always transformed to a QUBO \cite{mandal2020compressed}}. To this end, we write the canonical form of optimality and feasibility cuts constraints, adding scalar variables $a_{1}^{t}, a_{2}^{k} \geq 0$ as:
\begin{align}
\boldsymbol{p}^{t}(\mathbf{b} - \mathbf{A} \boldsymbol{z})+a_{1}^{t}==s \label{rand_1}\\
\boldsymbol{r}^{k}(\mathbf{b} - \mathbf{A} \boldsymbol{z})+a_{2}^{k}==0 \label{rand_2}
\end{align}
Then, following the table shown in \cite{zhao_quantum_milp}, we add the penalty weights $P_1, P_2$ and we raise the constraints to the square, before writing the resulting QUBO as:
\begin{align}
    \max_{ \boldsymbol{z} } \quad & \mathbf{i}^{T} \boldsymbol{z} + s + ||{\boldsymbol{z}}-{\boldsymbol{z}}^{(n-1)}|| + \tag{MP-QUBO} \label{master_qubo}\\
   & \sum_{ t \in T } P_{1} (\boldsymbol{p}^{t}(\mathbf{b} - \mathbf{A} \boldsymbol{z})-s+a_{1}^{t})^{2}+ \notag \\
   & \sum_{k \in K}P_{2} (\boldsymbol{r}^{k}(\mathbf{b} - \mathbf{A} \boldsymbol{z})+a_{2}^{k})^{2} \notag
\end{align}
However, \eqref{master_qubo} still consists of continuous variables, which cannot be understood by the QC. To tackle this problem we use a fundamental computer science technique, to approximate the continuous variables, called discretization or fixed-point approximation which is a two’s-complement number system that encodes positive and negative numbers in a binary representation as:
\begin{align*}
    s(\boldsymbol{p}) &\approx \sum_{i=0}^{a_{cc}}{2}^{i}\boldsymbol{p}^{pos}_{i}+\sum_{i=0}^{a_{cc}} 2^{-i} \boldsymbol{p}^{dec}_{i}-\sum_{i=0}^{a_{cc}} {2}^{i}\boldsymbol{p}^{neg}_{i} \\
    a_{1}^{t} &\approx \sum_{i=0}^{e^{1}_{cc}} 2^{i} \boldsymbol{a}_{i}^{[t]} \\
    a_{2}^{k} &\approx \sum_{i=0}^{e^{2}_{cc}} 2^{i} \boldsymbol{a}_{i}^{[k]}
\end{align*}
where $a_{cc}$ are the number of qubits that are used to approximate either the positive, decimal, or negative part of $s$. Furthermore, $t$ is the number of extreme points for $a_{1}^{t}$ and $k$ is the number of extreme rays. Notice that $s$ is approximated with some negative, positive, and decimal numbers, but $a_{1}^{t}, a_{2}^{k}$ are only approximated as positive integer variables. This does not degrade the quality of the solutions, because of the unconstrained nature of the resulting problem. The number of qubits required for approximating s is expressed by $a_{cc}$ which, must follow the following equation:
\begin{align} \label{qubits_start}
    a_{cc} \geq &\log_{2}(1+\max_{\boldsymbol{y}}{\boldsymbol{c}^T \boldsymbol{y}})
\end{align}
\hl{According to \eqref{qubits_start}, the number of qubits used to approximate the scalar variable $s$, which represents the contribution of the continuous function $\mathbf{c}^{T} \boldsymbol{y}$ to \eqref{master_optimized}'s objective function, must be sufficiently large to be greater than equal than the maximum value this function can take.} Moreover, following \eqref{qubits_start}:
\begin{align} 
    e^{1}_{cc} &\geq \log_{2}(\max_{\boldsymbol{z},s} {\boldsymbol{p}^{t}(\mathbf{b} - \mathbf{A} \boldsymbol{z})-s}) \\
    e^{2}_{cc} &\geq \log_{2}(\max_{\boldsymbol{z}} {\boldsymbol{r}^{k}(\mathbf{b} - \mathbf{A} \boldsymbol{z}}))  \label{qubits_start2}
\end{align}
\hl{Note that the auxiliary variables $a_{1}^{t}$ and $a_{2}^{k}$ need to be allocated with a sufficient number of qubits to accurately approximate the maximum value of the sum of the remaining terms in equations \eqref{rand_1} and \eqref{rand_2}, respectively.}
\begin{algorithm} 
\caption{Optimization \hl{Method} II} \label{alg:opt_group_2}
\KwData{$\text{Problem } \eqref{first_form}, R$}
\KwResult{$\boldsymbol{z}^{*},\boldsymbol{y}^{*}$}
$n \leftarrow 1$\;
$\boldsymbol{z},\overline{\boldsymbol{z}}^{(n)} \leftarrow 0$\;
\While{$|UB-LB|>\epsilon$}{
    \For{$\text{i $\leftarrow$ 0 to R}$}{
    
    $\text{Solve } \eqref{dualsub_sub_problem} \text{ and get } x_{i}^{(n)}({\boldsymbol{z}}), \boldsymbol{\lambda} $\;

     $LB \leftarrow x_{i}^{(n)}({\boldsymbol{z}}) $\;
     \If{$|UB-LB|>\epsilon$}{
        $\text{Finish}$\;
      }
     \eIf{$x({\boldsymbol{z}}^{(n)}))=-\infty$}{
        $\text{Add extreme ray } K \cup \{k\}$\;
      }{
        $\text{Add extreme point } T \cup \{t\}$\;
      }
      }
     $\text{Solve } \eqref{master_optimized} \text{ and get } s(\boldsymbol{p}),\boldsymbol{z}^{(n)}$\;
    $UB \leftarrow s(\boldsymbol{p}) $\;
    $n \leftarrow n+1$\;
}
\end{algorithm}

\hl{We note that according to \cite{doi:10.1287/mnsc.20.5.822}, it is feasible to ensure convergence without the necessity of solving the MP to optimality in each iteration. However, this study does not delve into the practical implications of such an approach.}
\section{Benchmark Models} \label{benchmarks}
In this section, we outline the formulation of both the OTS and NN verification problem. We have selected these two problems as benchmarks because the former's solution is pivotal in Power Systems, while the latter presents a relatively novel problem with the potential to transform Power Systems Operation.
\subsection{Optimal Transmission Switching (OTS)} \label{OTS}
The Optimal Transmission Switching (OTS) is a fundamental power systems problem. It involves maximizing the efficiency of the current system and addressing increasing demand using the already established infrastructure. The model of OTS is a mixture of an Optimal Power Flow (OPF), alongside binary decisions for the optimal switching of transmission lines. Therefore, we can formulate the OTS problem as:
\begin{subequations}
    \begin{align}
       \min_{g_{i}, p_{l}, \theta_{i},x_{l}} & \sum_{i \in \mathcal{B}} c_{i} g_{i} \\
                                        s.t. &\keno |p_{l}| \leq \overline{P}_l x_{l} \keno \forall l \in \mathcal{L}\\
                                             &0 \leq g_{i} \leq \overline{P}_i \\
                                             &g_{i}+ \sum_{l \in \mathcal{L}_{i}^{+}}p_{l} x_{l} = \sum_{l \in \mathcal{L}_{i}^{-}}p_{l} x_{l}+ {D}_i\\
                                             &p_{l} \geq B_{i,j} (\theta_i-\theta_j) - M (1-x_{l})  \\
                                             &p_{l} \leq B_{i,j} (\theta_i-\theta_j) + M (1-x_{l}) \\
                                             &\sum_{l \in \mathcal{L}} (1-x_{l}) \leq E \label{parameter_E} \\
                                             & \hl{g_{i}, \theta_{i} \in \mathbb{R} \keno \forall i,j} \\
                                             & \hl{x_{l} \in \{0,1\}, p_{l} \in \mathbb{R} \keno \forall l \in \mathcal{L}}
    \end{align}
\end{subequations}
In this formulation, we have used the DC-OPF approximation as it serves our purposes. $g_{i}$ expresses the generation dispatch of generators in node $i$ and $p_{l}$ expresses the power flow in line $l$, while $\overline{P}_l$ is the maximum power limit of the line. Furthermore, $\mathcal{L}_{i}^{+}$ is the set of lines that inject power to the node $i$, and $\mathcal{L}_{i}^{-}$ is the set of lines that export power from the node. Moreover, ${D}_i$ represents the demand in bus $i$. Finally, $\theta_i$ represents the voltage magnitude in bus $i$,  $B_{i,j}$ is the susceptance of line $l=(i,j)$ and $x_{l}$ is the variable that denotes the status of the line i.e. for $x_l=1$ the line is operational and for $x_l=0$ the line does not transmit power. 

\subsection{Neural Network verification for power systems} \label{NN_verification}
Neural Network constraint violation verification is a crucial step toward finding a scalable and trustworthy parametrized policy for problems in safety-critical systems. Recently, in \cite{venzke2020learning}, the authors proposed a direct MILP transformation of Neural networks that ensures precise worst-case performance guarantees, on constraints violations, across the entire range of possible inputs, for the DC-OPF problem:
\begin{subequations}
\begin{align}
    \max_{\hl{\mathbf{\hat{z}}_{k},\boldsymbol{z}_k ,\pd,\boldsymbol{\hat{p}}_{g}}}& p_g - \overline{p}_g \text{ or } \underline{p}_g-p_g\\
    & s.t. \quad \minpd \leq \pd \leq \maxpd \\
    & \pgslack = \sum_{i \in \setbuses} (\mathbf{M}_{d} \pd)^{i} - \sum_{i \in \setbuses \backslash \text{slack}} (\boldsymbol{\hat{p}}_{g}^{\text{nsg}})^i \label{opf}\\
    & \mathbf{\hat{z}}_{1}=\mathbf{{W}}_{1} \pd + \mathbf{{b}}_{1} \label{start_NN}\\
    & \mathbf{\hat{z}}_{k+1}=\weightk \mathbf{{z}}_{k} + \biask \label{eq24}\\
    & \boldsymbol{\hat{p}}_{g}^{\text{nsg}}=\weightK \mathbf{{z}}_{K} + \biasK \label{eq25}\\
    & \boldsymbol{z}_k = \boldsymbol{\sigma}({\boldsymbol{\hat{z}}_{k}})  \\
    &\hl{\mathbf{\hat{z}}_{k},\boldsymbol{z}_k  \in \mathbb{R} \keno \forall k} \\
    & \hl{\pd \in \mathbb{R} \keno \forall d} \\
    &{\hat{\boldsymbol{p}}_{g} \in \mathbb{R} \keno \forall g} \label{end_NN}
\end{align}
\end{subequations}
Constraints \eqref{start_NN}-\eqref{end_NN} are the exact representation of a Neural Network with $K$ layers, and neurons with ReLU activation functions. In \eqref{end_NN}, $\boldsymbol{z}_k$ expresses the output vector of neurons in layer $k$ while $\boldsymbol{\hat{z}}_{k}$ expresses their input vector. Function $\sigma(.)$ is the activation function, applied elementwise on vector $\boldsymbol{\hat{z}}_k$. In this paper, and in most neural network implementations in recent years, we use the ReLU activation function, which can be expressed as $\max(\boldsymbol{\hat{z}}_k, 0)$. Following \cite{venzke2020learning}, we use the big-M method to convert this to a set of linear constraints with binaries (for more information, please see \cite{venzke2020learning}). In \eqref{eq24}-\eqref{eq25}, Matrix $W$ represents the linear weights between layers $K$ and $K+1$, and vector $\boldsymbol{b}_{k+1}$ represents the biases in each neuron of layer $K+1$, which are added before the application of the activation function $\sigma(.)$ in layer $K+1$. Following this work we integrated in the aforementioned formulation the DC-OPF equations \eqref{opf}. In particular, variables $\pd$ and ${\hat{\boldsymbol{p}}}_{g}^{nsg}$ are the vectors of demand and generation (without the slack bus generation) respectively. Consequently, by changing the objective function, one could determine the maximum constraint violation of the trained NN, for any DC-OPF constraint. \hl{However, in the current formulation, the objective is to find the maximum upper or lower limit violation of the active power generation $p_g \in {\hat{\boldsymbol{p}}}_{g}$ of the generator $g$, where ${\hat{\boldsymbol{p}}}_{g}:=\{{\hat{\boldsymbol{p}}}_{g}^{nsg},\pgslack \} $}


A major drawback of this NN verification method is its scalability, although it is very accurate. Specifically, since Neural Networks can become very large, it is very hard computationally to run the resulting verification problems. The proposed framework could, theoretically,  reduce the computational burden of this problem, in ideal settings.

\begin{figure}[t!]
    \centering
    \includegraphics[width=\columnwidth]{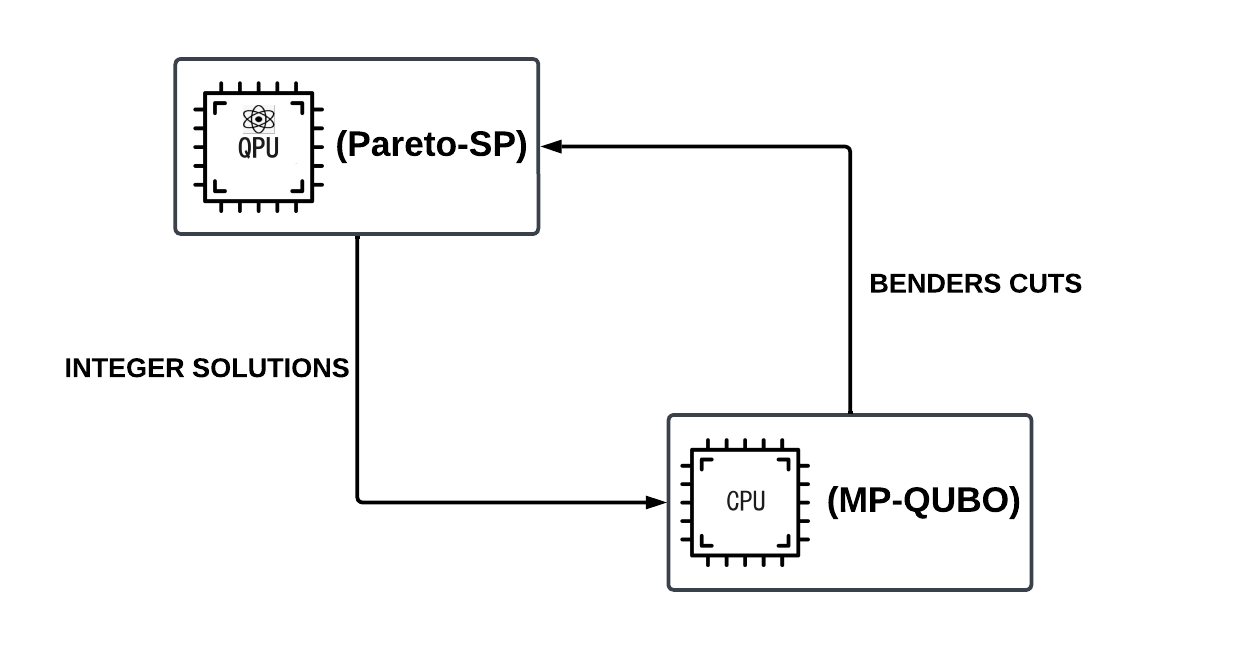}
    \caption{Solution Workflow of Proposed Methodology}
    \label{fig:flowchart}
\end{figure}

\section{Numerical Evaluation} \label{numericals}
In this section, we present the numerical evaluation of our method, validating the proposed framework. For the quantum solution we used the Hybrid solver provided by D-Wave Leap \cite{DWaveLeap} and for the classical solver we used Gurobi \cite{gurobi}. \hl{The solution workflow, showing the information exchange between CPU and QPU can be shown in Fig.\ref{fig:flowchart}.} 
We denote that for "Mean Iteration Time" in Table \ref{Benchmarks_OTS_6}, Table \ref{Benchmarks_OTS} and Table \ref{Benchmarks_NN}, we include only CPU and QPU Access time \hl{\cite{Operation}}. Moreover, with "Minimum objective function until Iteration" on the y axis of Fig.\ref{fig:c}-\ref{fig:h}, we mean $min(UB)$ and $max(LB)$ found until a given iteration in the x axis.

In the following section, we will compare the proposed methodology with the following:
\begin{itemize}
    \item Single Step MILP Optimization as shown in \eqref{first_form} (SSO)
    \item Conventional Benders Decomposed problem solved by CPUs only (C-BD-C)
    \item Accelerated Benders with optimization method I Decomposed problem solved by CPUs only (BD-C-I)
    \item Accelerated Benders with optimization method I Decomposed problem solved by CPU and QPU (BD-QC-I)
    \item Accelerated Benders with optimization method II Decomposed problem solved by CPU and QPU running the \eqref{dualsub_sub_problem}, R times per Benders Iteration (BD-QC-II-R). We provide numerical results for R=2,3,5,10
\end{itemize}
\hl{We note that the Quantum Annealing solutions used in this section and Quantum Approximate Optimization Solutions benchmarked in the next section, are heuristic solutions. Therefore the results of the Branch and Bound Algorithm of Gurobi for the solution of (SSO) are used as a ground truth, to assess the quality of the benchmarked solutions. }
\subsection{Benders Benchmarks}
\begin{figure}[t!] 

\begin{subfigure}{0.24\textwidth}
\includegraphics[width=\linewidth]{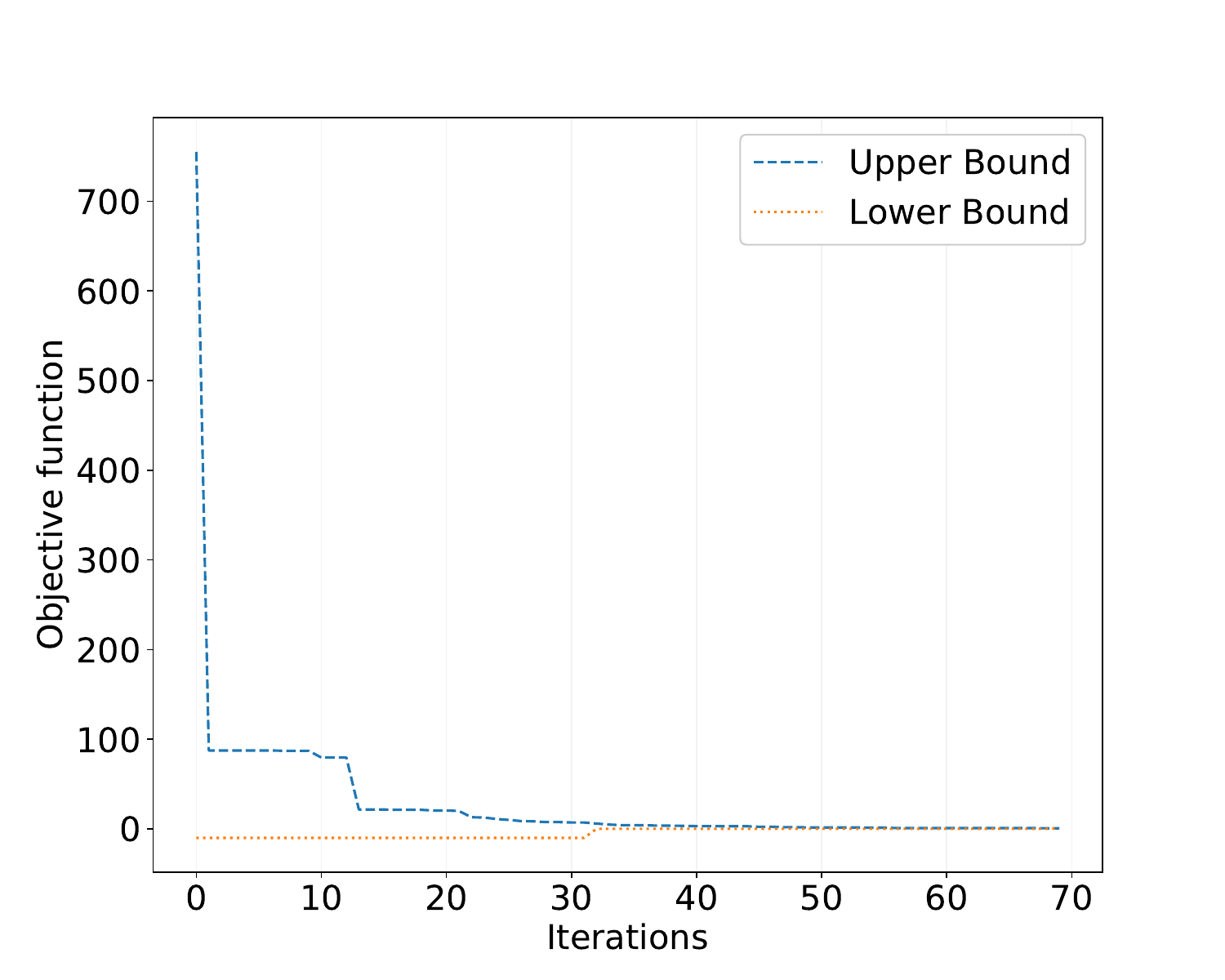}
\caption{(BD-C-I) solving NN verification} \label{fig:a}
\end{subfigure}\hspace*{\fill}
\begin{subfigure}{0.24\textwidth}
\includegraphics[width=\linewidth]{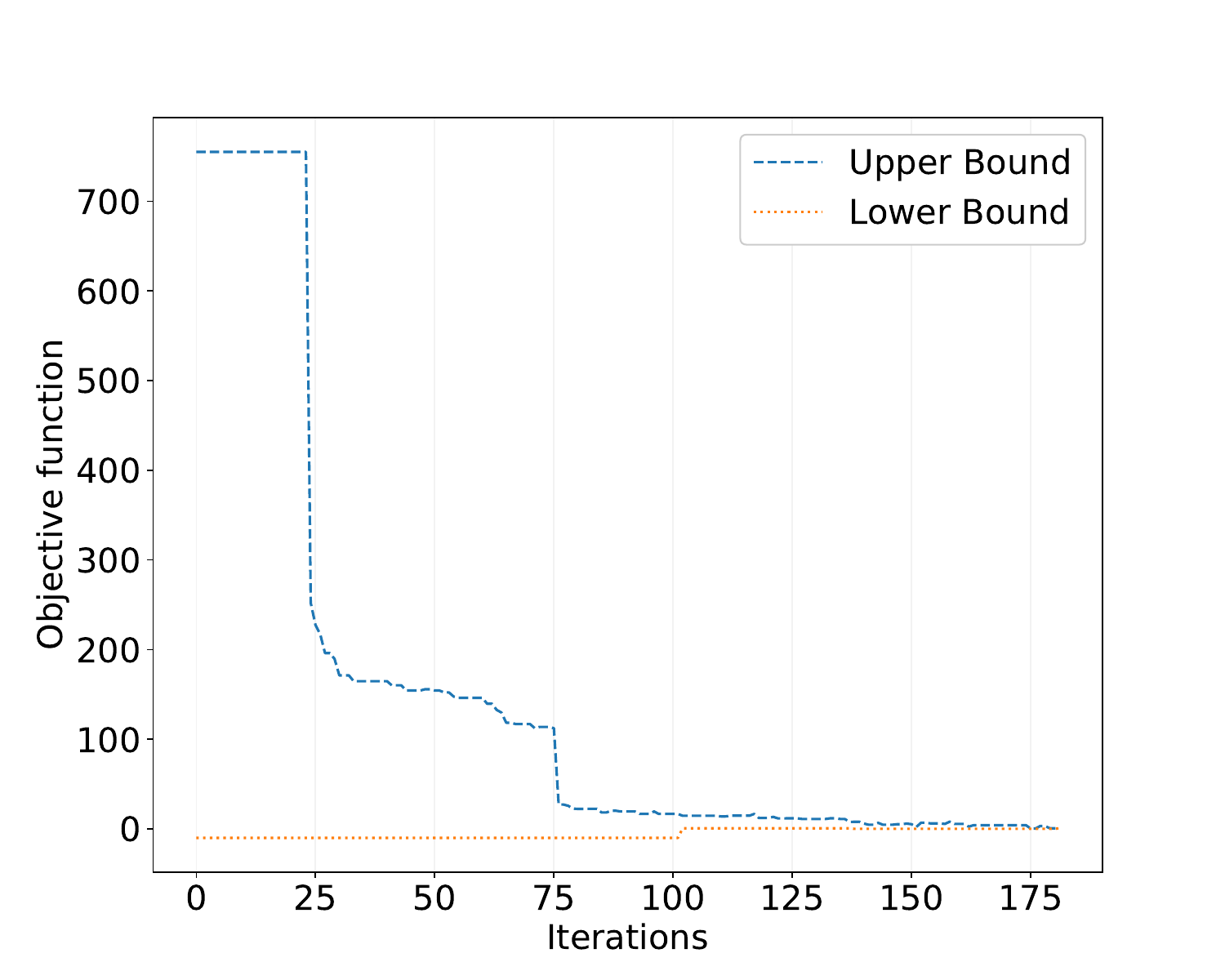}
\caption{(C-BD-C) solving NN verification} \label{fig:b}
\end{subfigure}
\begin{subfigure}{0.24\textwidth}
\includegraphics[width=\linewidth]{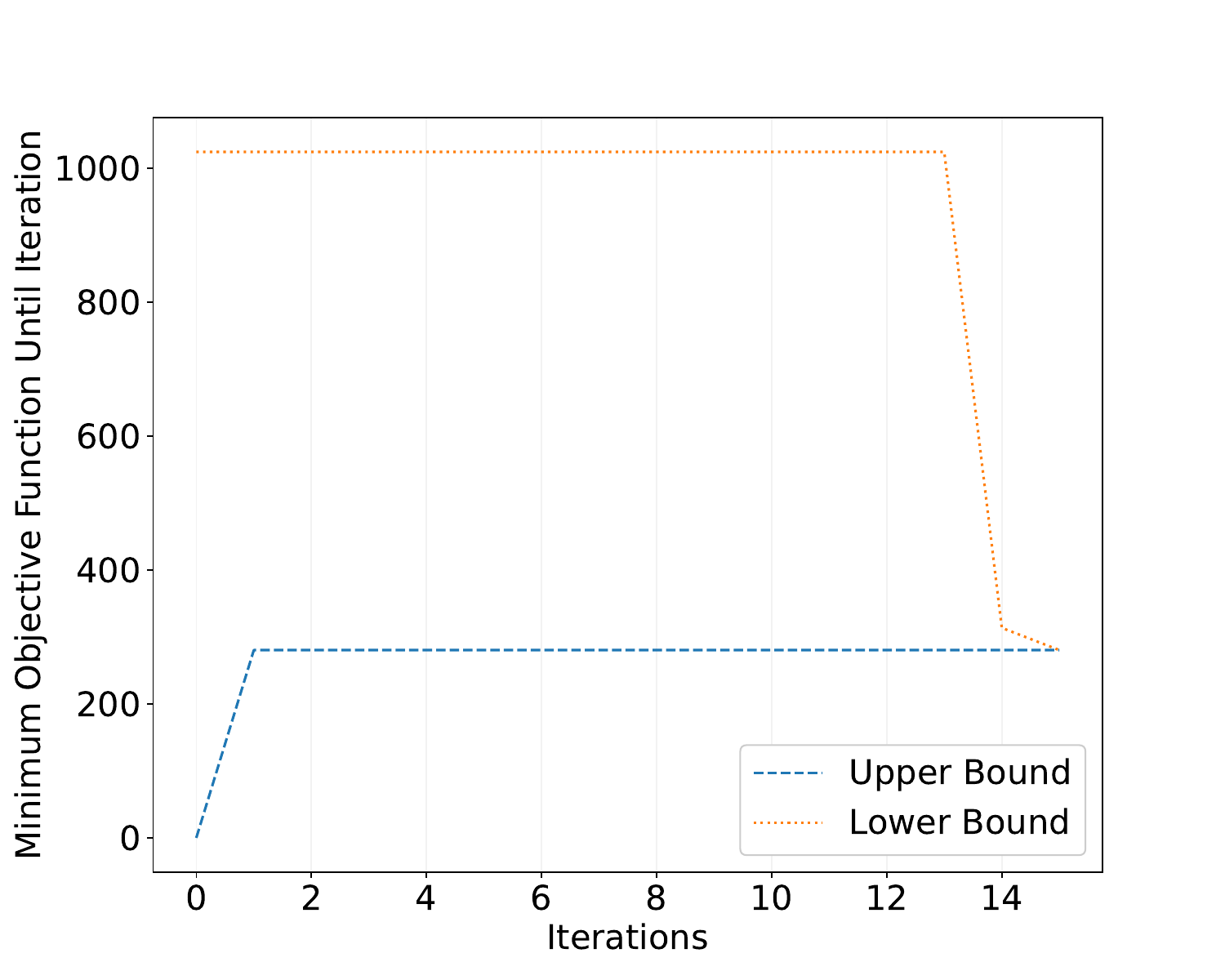}
\caption{(BD-C-I) solving OTS} \label{fig:c}
\end{subfigure}\hspace*{\fill}
\begin{subfigure}{0.24\textwidth}
\includegraphics[width=\linewidth]{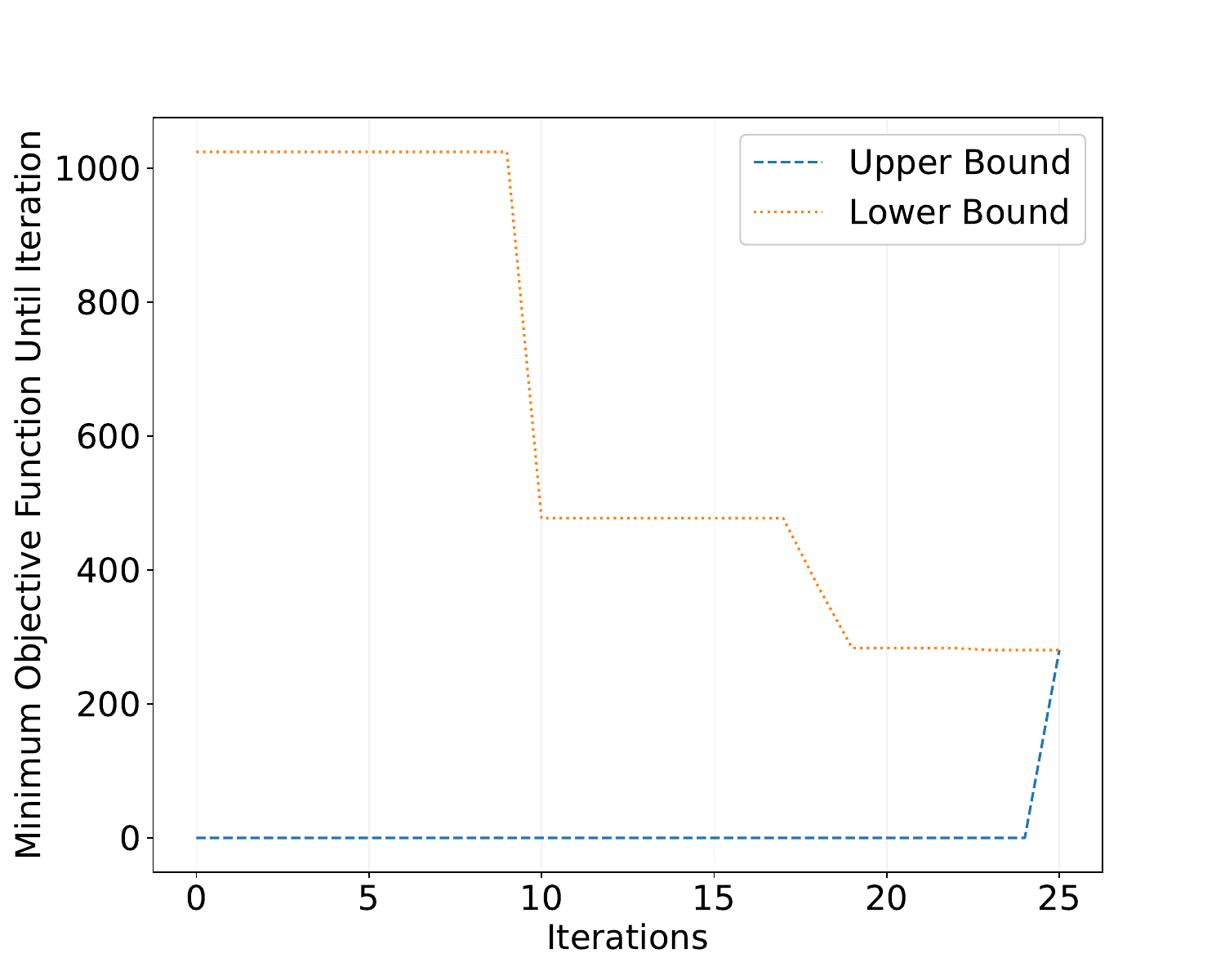}
\caption{(C-BD-C) solving OTS} \label{fig:d}
\end{subfigure}
\begin{subfigure}{0.24\textwidth}
\includegraphics[width=\linewidth]{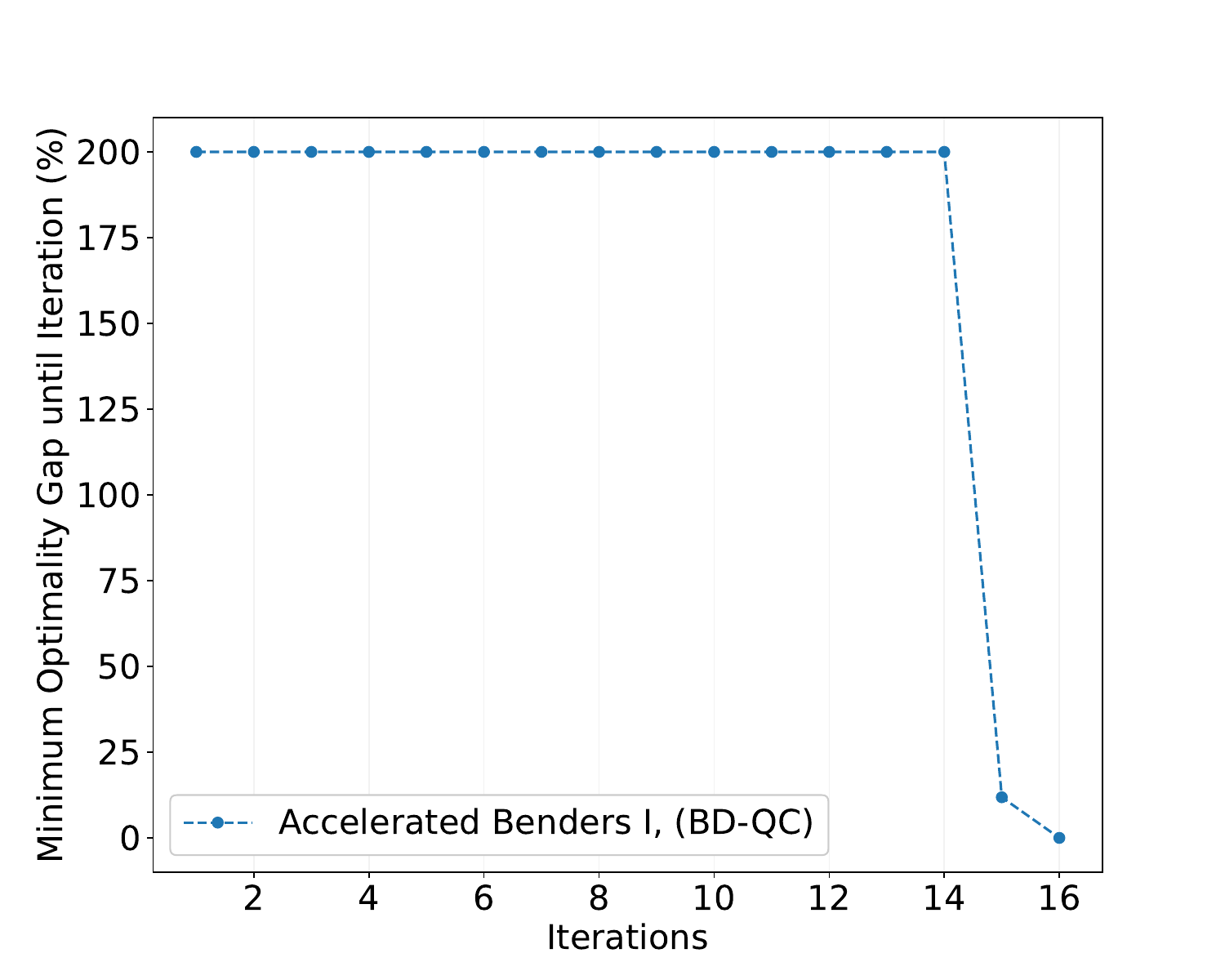}
\caption{(BD-QC-I) solving OTS for IEEE 14 Bus} \label{fig:e}
\end{subfigure}\hspace*{\fill}
\begin{subfigure}{0.24\textwidth}
\includegraphics[width=\linewidth]{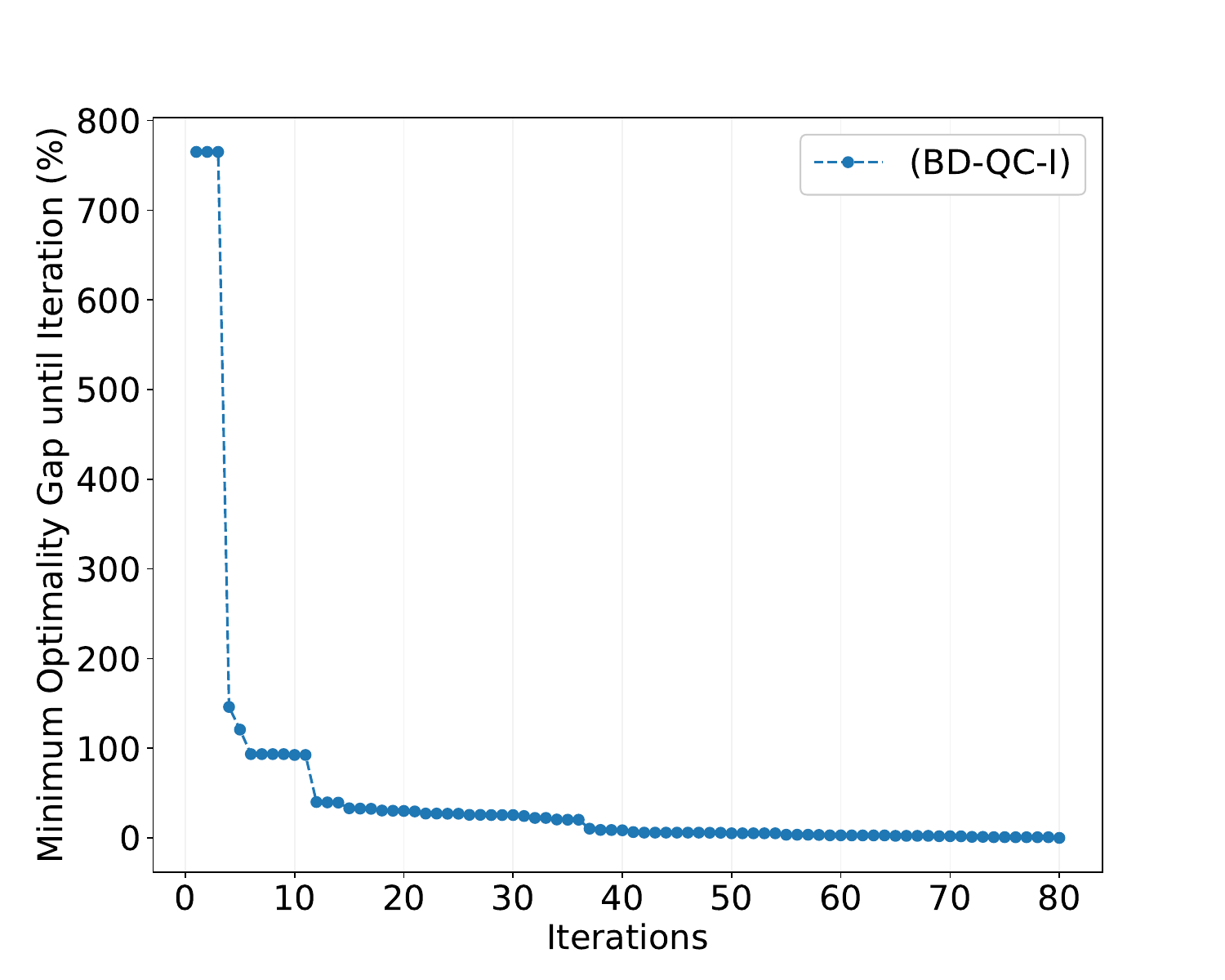}
\caption{(BD-QC-I) solving NN Verification} \label{fig:f}
\end{subfigure}
\begin{subfigure}{0.24\textwidth}
\includegraphics[width=\linewidth]{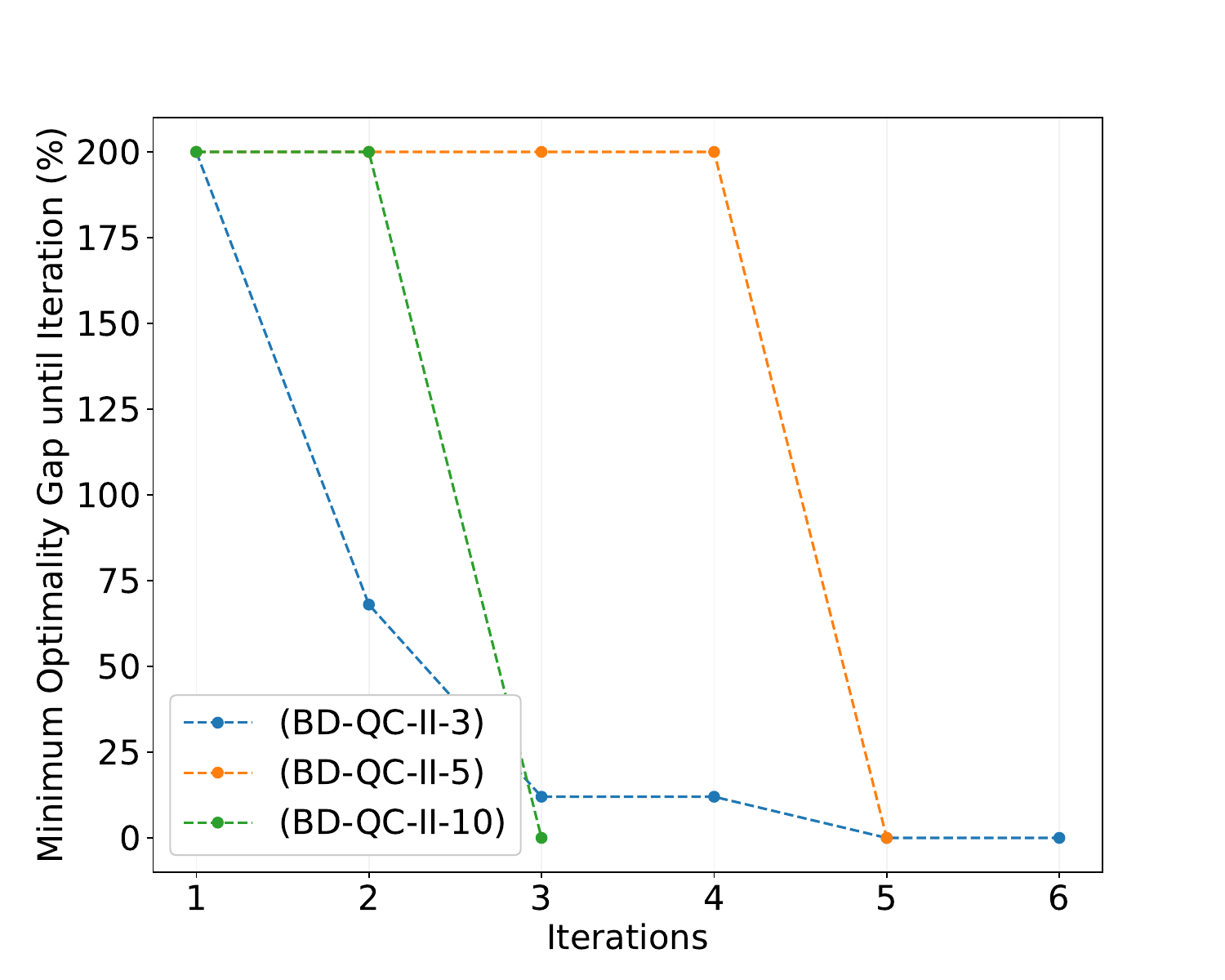}
\caption{(BD-QC-II) solving OTS for IEEE 14 Bus} \label{fig:g}
\end{subfigure}\hspace*{\fill}
\begin{subfigure}{0.24\textwidth}
\includegraphics[width=\linewidth]{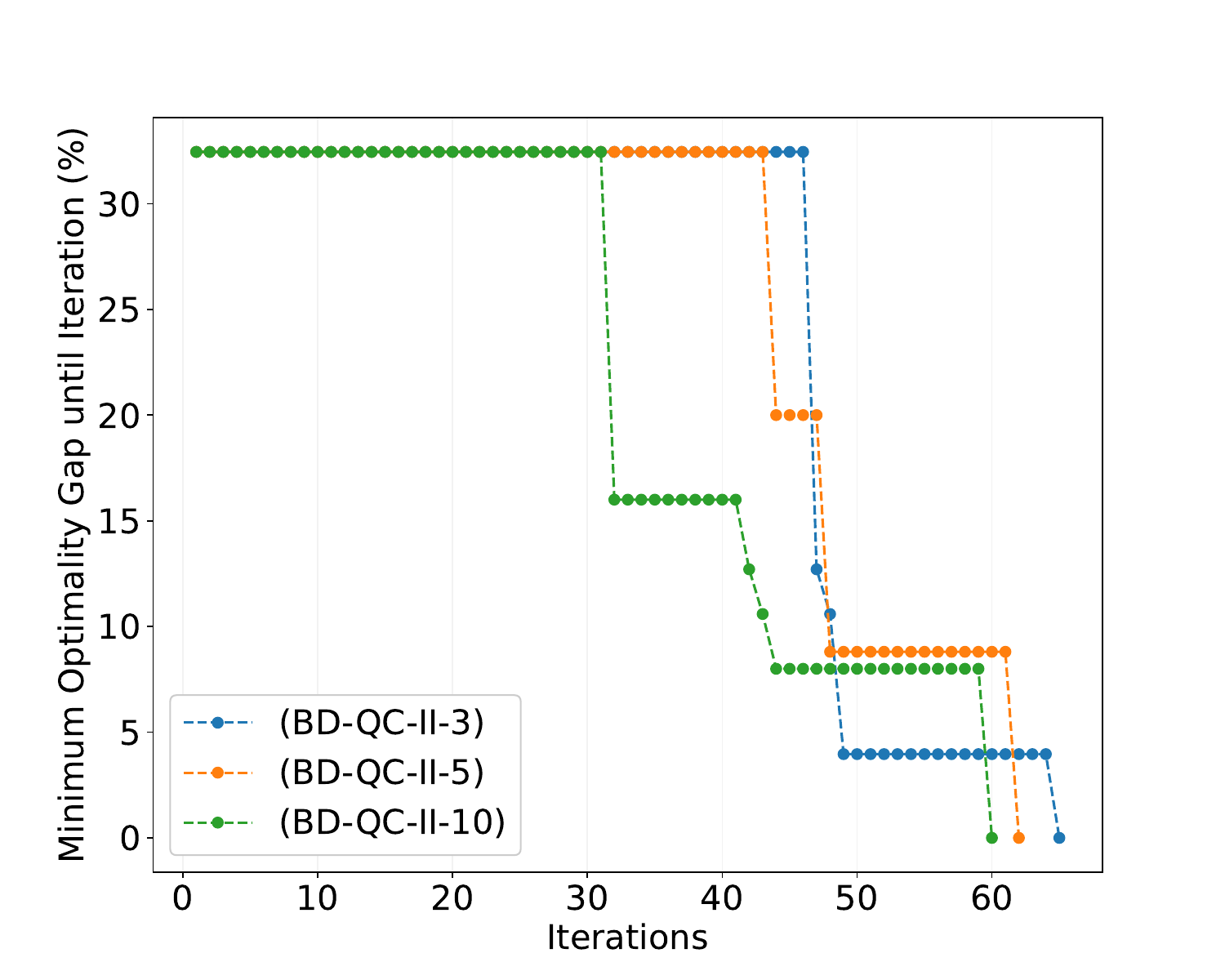}
\caption{(BD-QC-II) solving NN verification} \label{fig:h}
\end{subfigure}
\caption{} \label{fig:four_graphs}
\end{figure}

Initially, we assess the efficacy of the enhancements introduced in comparison to the conventional Benders decomposition method. Therefore, we benchmark the conventional Benders decomposition (C-BD-C) against the proposed accelerated versions (BD-C-I). Figures\ref{fig:a} and \ref{fig:b}, illustrate the disparity in convergence between the accelerated (BD-C-I) and the conventional (C-BD-C) approaches for the NN verification problem and the OTS problem, respectively. The results demonstrate the substantial positive influence of the algorithmic modifications on both performance and convergence speed. Specifically, for NN verification, the proposed enhancements accelerated the algorithm by 61\%, and in the case of OTS, the acceleration was 38\%. Furthermore, it is noteworthy that the trailing effect was significantly mitigated in the BD-C-I approach.

Moreover, comparing the iterations needed for (BD-QC-II-R) with (BD-QC-I) we can view that for both NN verification and OTS there was a reduction in the number of iterations needed to converge. This happens because in every iteration we produce more cuts. However, as Tables \ref{Benchmarks_OTS_6},\ref{Benchmarks_OTS} and \ref{Benchmarks_NN} show, in the case of NN verification (BD-QC-II-R) methodology does worse than (BD-QC-I) in all cases, as every iteration of the former needs more time than every iteration in the latter.

\subsection{OTS Benchmarks} \label{OTS_bench_sec}
\subsubsection{IEEE 6 Bus Benchmark} 
Initially, we conducted experiments using the IEEE 6-Bus system \cite{ieeebus6}, which consists of 11 lines, to assess the effectiveness of two methodologies, (BD-QC-I) and (BD-QC-II-2), in comparison to the established benchmarks of (C-BD-C) and (SSO). For this evaluation, we executed each benchmark five times while setting the parameter $E$ in \eqref{parameter_E} to 5. \hl{Furthermore, the parameters $a_{cc}, e^{1}_{cc}, e^{2}_{cc}$ were set to 10.} The average solution times for these instances are presented in Table \ref{Benchmarks_OTS_6}.

Notably, it is evident that (BD-QC-II-2) outperforms the other methodologies, indicating a clear quantum advantage in this context. Additionally, (BD-QC-I) appears to be superior to the conventional approach represented by (C-BD-C), but inferior of (SSO).
\begin{table}[h]
\centering
\caption{\hl{Benchmarks comparison for the Optimal Transmission Switching problem for the IEEE 6-Bus system}} \label{Benchmarks_OTS_6}
\begin{tabular}{ |p{1.7cm}||p{1.7cm}|p{1.5cm}|p{1.5cm}|  }
 \hline
 \multicolumn{4}{|c|}{\hl{6-Bus} OTS Results} \\
 \hline
 Method&  Mean Iteration Time (s) &  Iterations & Objective Value (MW)\\
 \hline
 SSO   &0.281 &0&   280.32\\
 C-BD-C& 0.043 & 8   &280.32\\
 BD-QC-I &0.050 & 7&  280.32\\
 BD-QC-II-2 &0.056& 3 &  280.32\\
 \hline
\end{tabular}
\end{table}

\subsubsection{IEEE 14 Bus Benchmark} 
Assessing the scalability of the hybrid classical-quantum algorithms, we expand our numerical results, by solving larger instances of OTS, on the IEEE 14 Bus system. This evaluation involved optimizing 20 binary variables, with the parameter $E$ set to 17. \hl{Moreover, the parameters $a_{cc}, e^{1}_{cc}, e^{2}_{cc}$ were set to 18.} Table \ref{Benchmarks_OTS} presents the results for the OTS solutions on the 14-bus system. It is essential to highlight that all methods successfully achieved the optimal value. What is particularly striking is that some methods relying on harnessing QCs gave close results to those relying solely on conventional CPUs. In this regard, (SSO) emerges as a standout solution. Intriguingly, (BD-QC-II-5) outperforms the other benchmarks except (SSO), which has a comparable solution time. It is worth noting that, (BD-C-II-5) outperforms its counterparts, because the others either result in unnecessary cuts or result in too many QPU calls. Therefore, parameter $R$ became a tunable hyperparameter. 

As shown in Fig.\ref{fig:c} and Fig.\ref{fig:e} (BD-QC-I) takes one iteration more than (BD-C-I), which can be possibly attributed to the inherent noise associated with NISQ. Finally, in Fig.\ref{fig:g}, the behavior of the methods (BD-QC-II) is depicted.


\begin{table}[h]
\centering
\caption{\hl{Benchmarks comparison for the Optimal Transmission Switching problem for the IEEE 14-Bus system}} \label{Benchmarks_OTS}
\begin{tabular}{ |p{1.7cm}||p{1.7cm}|p{1.5cm}|p{1.5cm}|  }
 \hline
 \multicolumn{4}{|c|}{\hl{14-Bus} OTS Results} \\
 \hline
 Method&  Mean Iteration Time (s)  &  Iterations & Objective Value (MW)\\
 \hline
 SSO   &0.38 &0&   280.32\\
 BD-C-I&   0.04  & 15   &280.32\\
 BD-QC-I &0.05& 16&  280.32\\
 BD-QC-II-3 &0.07& 6&  280.32\\
 BD-QC-II-5 &0.08& 5&  280.32\\
 BD-QC-II-10 &0.28& 3&  280.32\\
 \hline
\end{tabular}
\end{table}

\subsection{NN Verification Benchmarks}
Neural Network verification was more challenging to solve, as we can see from the increased number of iterations needed to converge. The tested neural networks contained 3 input neurons, and 3 output neurons alongside 2 hidden layers with 20 neurons each.  The NN was trained to determine the DC-OPF result applied on the IEEE 9-bus system. \hl{Parameters $a_{cc}, e^{1}_{cc}, e^{2}_{cc}$ were set to 20.}
Moreover,  In Fig.\ref{fig:f} we can observe that the algorithm converges in 79 iterations, in contrast with (BD-C-I) method which only needed 70 iterations to converge. Similar to the previous test case in section \ref{OTS_bench_sec}, (SSO) was the best benchmark. In this instance, we can observe that (BD-QC-I) outperforms all (BD-QC-II) methods, in solution time. It becomes, thus, obvious that different decomposition techniques are most suitable for delivering the best results in different problems. Finally, parameter $R$ is crucial for the performance enhancement of the method (BD-QC-II).

\begin{table}[h]
\centering
\caption{\hl{Benchmarks comparison for the NN Verification problem}} \label{Benchmarks_NN}
\begin{tabular}{ |p{1.7cm}||p{1.7cm}|p{1.5cm}|p{1.5cm}|  }
 \hline
 \multicolumn{4}{|c|}{NN Results} \\
 \hline
 Method&  Mean Iteration Time (s)  &  Iterations &Objective Value (MW)\\
 \hline
 SSO   &0.19 &0&   0.74\\
 BD-C-I&   0.05  & 69   &0.74\\
 BD-QC-I &0.06& 79&  0.74\\
 BD-QC-II-3 &0.09& 65&  0.74\\
 BD-QC-II-5 &0.12& 63&  0.74\\
 BD-QC-II-10 &0.16& 60&  0.74\\
 \hline
\end{tabular}
\end{table}

\section{Discussion}
\label{sec:discussion}
Besides our experiments with Quantum Annealers (such as D-Wave LEAP), which we presented in the previous Section, we also explored implementing the aforementioned simulations using circuit-based QCs by leveraging the Quantum Approximate Optimization Algorithm (QAOA). These experimental runs were conducted on IBM's latest available Eagle r3 Processor, with capacity of 127 qubits. Circuit-based QCs are considered general purpose QCs, and are expected to become one of the main quantum computing technologies in the future. In contrast, Quantum Annealers have been tailored to solve specifically optimization problems. Still, circuit-based QCs are less mature compared to the Quantum Annealers, when it comes to optimization problems, and suffer from increased average error per gate. Beyond a certain circuit depth, which is between 60-100 gates at the moment (October 2023), real circuit-based QCs become too noisy. The circuit depth required for our QAOA implementation to solve \eqref{master_qubo} was 500 gates, which hampered our efforts to obtain a solution. Hence, it appears that Quantum Annealing might be a more suitable alternative for solving unconstrained binary optimization problems at this stage. \hl{This claim is supported by the recent work \cite{Pelofske_2023}, which empirically proves that Quantum Annealing technologies will constantly perform better than circuit based QAOA for combinatorial optimization problems solution. } \hl{ Therefore, in this work, we showed that quantum annealing} can already deliver faster solutions for smaller power system sizes, and has the potential to scale to larger problems in the near future. At the same time, as a research community we must also concentrate our efforts in formulating optimization procedures with reduced circuit depth for circuit-based QCs. Considering the major efforts taking place in the quantum computing community, in the near future improved error mitigation strategies will appear which will address several of the current scalability challenges of quantum computing. The power systems community shall be ready with the suitable formulations for quantum optimization tools in order to directly exploit the potential for faster computing times. 

As far as the methods we presented in this paper are concerned, besides circuit depth, one additional key factor to consider when we design a quantum optimization method is how our chosen method affects the qubit requirements. As we use Benders decomposition method, the need for qubits increases linearly with the number of constraints in \eqref{master_optimized}. This happens because we introduce integer variables such as $a_{1}^{t}$ and $a_{2}^{k}$ in each constraint, which in turn affects the number of qubits required, as explained in \eqref{qubits_start} to \eqref{qubits_start2}. Developing approaches that could arrive at similar results with a reduced number of qubits will lead both to scalable solutions and faster computing times. To facilitate these efforts, we have placed the Python code for all the presented approaches online, in the form of a tutorial, so that the research community may use this as a starting point for future improvements.

\section{Conclusion}
\label{sec:conclusion}

This paper introduces a framework for transforming Mixed Integer Linear Programs (MILPs) for power system problems to a "quantum-classical solvable form" in order to exploit the benefits of quantum optimization solvers.
MILPs play a pivotal role in power system operation and planning, including e.g. unit commitment and optimal transmission switching, while they have recently emerged as a tool for trustworthy machine learning, e.g. for solving neural network verification problems for power systems. We introduce a general framework employing a hybrid “quantum-classical” approach and specifically utilizing an accelerated version of Benders decomposition. We apply this framework to two use cases: Optimal Transmission Switching and Neural Network verification for DC-OPF. We show that for small problem sizes, Quantum Annealers already achieve faster computation times than classical computing. Our experiments with larger problems in Quantum Annealers and real circuit-based QCs showed that future work needs to explore how to exploit the advantages of quantum optimization through scalable methods that require reduced circuit sizes. Therefore, to facilitate future efforts in the power systems research community, one of the contributions of this paper is that all our code, in the form of a tutorial in Python, is online available in \cite{github}. 


\bibliographystyle{ieeetr} 
\bibliography{biblio_paper}

\end{document}